\documentclass[nofootinbib,showpacs,preprintnumbers,amsmath,amssymb,floatfix]
{revtex4}
\usepackage{graphicx}
\usepackage{epsf}
\usepackage{wrapfig}
\begin{document}


\title{Unpolarized DIS structure functions in Double-L8ogarithmic Approximation}

\vspace*{0.3 cm}

\author{B.I.~Ermolaev}
\affiliation{Ioffe  Institute, 194021
 St.Petersburg, Russia}
\author{S.I.~Troyan}
\affiliation{St.Petersburg Institute of Nuclear Physics, 188300
Gatchina, Russia}

\begin{abstract}
We present description of the DIS structure functions $F_1$ and $F_2$ at small $x$ obtained in double-logarithmic approximation (DLA).
 First we clarify our previous results on $F_1$  and then obtain explicit expressions for $F_2$. Our calculations confirm our previous
 result that
the small-$x$ asymptotics of $F_1$ is controlled by a new Pomeron that has nothing to do
with  the BFKL Pomeron, though their intercepts are pretty close. The latter means that studying the small-$x$ dependence of the
unpolarized DIS cannot ascertain which of those Pomerons is actually involved. However, we predict a quite different and universal $Q^2$-dependence of $F_{1,2}$ in DLA
compared to the approaches involving the both DGLAP and BFKL. On that basis, we construct simple relations between logarithms
of $F_{1,2}$, which can be verified with analysis of experimental data.
In contrast to $F_1$, the intercept controlling the small-$x$ asymptotics of $F_2$ is very small
but positive, which ensures growth of $F_2$ at $x \to 0$.
\end{abstract}


\maketitle

\section{Introduction}

Theoretical investigation of  the structure functions $F_1$ and $F_2$ in the QCD framework includes both fixed-order
calculations\cite{flor}-\cite{moch3} and approaches operating with all-order resummations,
which
usually either are based on DGLAP\cite{dglap} (see e.g. Ref.~\cite{kotik2}) or combine DGLAP and BFKL\cite{bfkl}. In the latter case,
the role of BFKL is providing the structure functions
with the necessary growth at small $x$ while the role of DGLAP is describing the $Q^2$ dependence (see e.g.
Refs.~\cite{kwiech, ball}).
There also are more involved approaches engaging  the dipole model\cite{dipol}.
For example,
Ref.~\cite{kovch} involves contribution of the multiple Pomeron exchange to $F_2$. Numerical
analysis of HERA data within the dipole model can be found in
Ref.~\cite{lusz}. The dipole model
was used in the global analysis of experimental data in Ref.~\cite{arm}.
A detailed bibliography on this issue can be found in Ref.~\cite{zfitter}. Let us remind that growth  of $F_2$
at small $x$ was also suggested in the Regge inspired models\cite{cap,bart}
in the context of Diffractive DIS.

In contrast to the aforecited approaches, we suggest an alternative description of $F_{1,2}$, which is not based on DGLAP and BFKL as well as on
their modifications. Namely, we calculate $F_1$ and $F_2$ in Double-Logarithmic Approximation (DLA).
In order to calculate and sum DL contributions to $F_{1,2}$ to all orders in
$\alpha_s$, we construct and solve Infra-Red Evolution
Equations (IREE). This method was initiated by L.N.~Lipatov~\cite{kl} and then it
has been applied many times to a wide spectrum of calculations in QCD and Standard Model. The basis of this method
was constructed in the pioneer works\cite{ggfl} where it was found that DL contributions come equally from
virtual
quarks and gluons with small transverse momenta.
In the context of the Unpolarized DIS, the IREE method was applied to calculating $F_1$ in Ref.~\cite{etf1} and $F_L$ in Ref.~\cite{etfl}.
Applications of IREE to the high-energy spin physics can be found in Ref.~\cite{egtg1sum}.

Our paper is organized as follows: In sect.~II we introduce our definitions for invariant amplitudes which
simplify the structure of the hadronic tensor. In Sect.~III we compose IREEs for $F_1$ and $F_2$. Actually, IREEs for
$F_1$ were already  constructed and solved in Ref.~\cite{etf1} but a part of DL contributions was not accounted for there.
This flaw is corrected in Sect.~III. Solutions to the IREEs are obtained in Sect.~IV. Problem of inputs to the IREEs
is considered in Sect.~V. After the inputs have been specified, the explicit expressions for $F_{1,2}$ were
obtained. The small-$x$ asymptotics of $F_{1,2}$ are obtained in Sect.~VI and they are used to derive simple relations between logarithms
of $F_{1,2}$. Finally, Sect.~VII is for concluding remarks.

\section{Invariant amplitudes simplifying calculation of  the hadronic tensor}

The unpolarized part of the hadronic tensor $W_{\mu\nu}$ describing DIS off a hadron is conventionally
parameterized as follows:

\begin{eqnarray}\label{wgen}
W_{\mu\nu} (p,q) &=& \left(g_{\mu\nu} - \frac{q_{\mu}q_{\nu}}{q^2}\right) F_1 + \frac{1}{pq}
\left(p_{\mu} - q_{\mu}\frac{pq}{q^2}\right)\left(p_{\nu} - q_{\nu}\frac{pq}{q^2}\right) F_2
\end{eqnarray}
The standard notations $q$ and $p$ in Eq.~(\ref{wgen}) stand for momenta of the virtual photon and hadron
respectively.
%
%
In  the framework of QCD factorization $W_{\mu\nu}$ can be represented through convolutions
of perturbative and non-perturbative contributions:

\begin{equation}\label{fact}
W_{\mu\nu} = W^{(q)}_{\mu\nu} \otimes \Phi_q +   W^{(g)}_{\mu\nu} \otimes \Phi_g,
\end{equation}
with $p$ standing for the initial parton momenta.
 $ \Phi_q $ and $ \Phi_g $ In Eq.~(\ref{fact}) are the initial quark and gluon distributions in the hadron respectively.
They accommodate non-perturbative contributions. In contrast, $W^{(q)}_{\mu\nu}$ and $W^{(g)}_{\mu\nu}$
are perturbative objects. They describe DIS off a quark (gluon) respectively. Parametrization of their tensor
structure
is similar  to Eq.~(\ref{wgen}):

\begin{eqnarray}\label{wqg}
W^{(q)}_{\mu\nu} (p,q) &=& \left(-g_{\mu\nu}
+ \frac{q_{\mu}q_{\nu}}{q^2}\right) F^{(q)}_1 + \frac{1}{pq}
\left(p_{\mu} - q_{\mu}\frac{pq}{q^2}\right)\left(p_{\nu} - q_{\nu}\frac{pq}{q^2}\right) F^{(q)}_2,
\\ \nonumber
W^{(g)}_{\mu\nu} (p,q) &=& \left(-g_{\mu\nu} + \frac{q_{\mu}q_{\nu}}{q^2}\right) F^{(g)}_1 + \frac{1}{pq}
\left(p_{\mu} - q_{\mu}\frac{pq}{q^2}\right)\left(p_{\nu} - q_{\nu}\frac{pq}{q^2}\right) F^{(g)}_2,
\end{eqnarray}
with $p$ being the initial parton momentum.
$F^{(q)}_{1,2}$ and $F^{(g)}_{1,2}$ in Eq.~(\ref{wqg}) are calculated by means of perturbative QCD.
The standard way is to calculate $F^{(q,g)}_{1,2}$ through auxiliary invariant amplitudes $A^{(q,g)},B^{(q,g)}$
defined as follows:

\begin{eqnarray}\label{defa}
A^{(q,g)} &\equiv& g_{\mu\nu} W^{(q,g)}_{\mu\nu} =
- 3 F^{(q,g)}_1 + \frac{F^{(q,g)}_2}{2x}  + O(p^2),
\end{eqnarray}

\begin{eqnarray}\label{defb}
B^{(q,g)} \equiv \frac{p_{\mu}p_{\nu}}{pq} W^{(q,g)}_{\mu\nu} =
-\frac{1}{2x} F^{(q,g)}_1 + \frac{1}{4x^2} F^{(q,g)}_2 ,
\end{eqnarray}
where we have used the standard notations $x = Q^2/w$ $(Q^2 = - q^2 >0)$, $w = 2pq$.  Eqs.~(\ref{defa},\ref{defb}) yield that

\begin{eqnarray}\label{fab}
F^{(q,g)}_1 &=& - \frac{A^{(q,g)}}{2} + x B^{(q,g)},
\\ \nonumber
F^{(q,g)}_2 &=& - x A^{(q,g)} + 6x^2B^{(q,g)} =
2xF^{(q,g)}_1 + 4x^2 B^{(q,g)}.
\end{eqnarray}

The longitudinal structure functions $F_L^{(q,g)}$ are related to $B^{(q,g)}$:

\begin{equation}\label{flqgb}
F_L^{(q,g)} = 4 x^2 B^{(q,g)}.
\end{equation}

Eq.~(\ref{fab}) reads that the contributions of $B^{(q,g)}$ to $F^{(q,g)}_{1,2}$ can
be neglected at small $x$ compared to $A^{(q,g)}$ only
if $ B^{(q,g)}$ is less singular than $x^{-1 - \sigma}$, with $\sigma > 0$.

\section{IREEs for the auxiliary amplitudes $A^{(q,g)}$ and $B^{(q,g)}$}

We calculate amplitudes $A^{(q,g)}$ and $B^{(q,g)}$ by constructing and solving IREEs for them.
The IREEs are differential, so the first step is to obtain general solutions to them. Then
the general solutions should be specified.
Up to this point the technology of treating amplitudes $A^{(q,g)}$ and $B^{(q,g)}$ is the same,
so we construct and solve IREEs for amplitudes $A^{(q,g)}$ and then extend our results to $B^{(q,g)}$. This
similarity  ends when specifying the general solutions begins, so at this point we will consider $A^{(q,g)}$ and $B^{(q,g)}$
separately. Throughout the paper we will use the Sudakov parametrization for virtual parton momenta $k_i$,
representing them as follows:

\begin{equation}\label{sud}
k_i = \alpha_i q' + \beta_i p' + k_{i\perp},
\end{equation}
where $q'$ and $p'$ are the massless (light-cone) momenta made of momenta $p$ and $q$:

\begin{equation}\label{pqprime}
p' = p - q (p^2/w) \approx p, ~~~ q^{\prime} = q - p (q^2/w) = q + x p.
\end{equation}

In Eq.~(\ref{pqprime}) $q$ denotes the virtual photon momentum while $p$ is momentum of the initial parton.
We remind that we presume that $p^2$ is small, so we will neglect it in what follows.

First step to construct IREEs is to introduce an artificial infrared cut-off $\mu$. It
is necessary procedure in DLA to regulate IR divergences. Since that amplitudes $A^{(q,g)}$ become
$\mu$-dependent, which makes possible to trace their  evolution with respect to $\mu$. In order to
relate this evolution to evolution in $x$ and $Q^2$, we parameterize
amplitudes $A^{(q,g)}$ as follows: $A^{(q,g)} = A^{(q,g)}\left(s/\mu^2, Q^2/\mu^2 \right)$.  Value of
$\mu$ is arbitrary save the obvious restriction
$\mu > \Lambda_{QCD}$ to enable applicability of the perturbative QCD. For instance, the factorization scale can be used as the
IR cut-off. As we are not interested in studying dependence of $A_{q,g}$ on masses $m_r$ of involved quarks and their virtualities,
we presume that $\mu > m_r$. So as to treat virtual quarks and gluons identically, we prescribe
The IREEs include convolutions of amplitudes, so it is convenient to write them in terms of the Mellin transform

\begin{eqnarray}\label{mellin}
 A_{q,g} \left(s/\mu^2, Q^2/\mu^2 \right) &=& \int_{- \imath \infty}^{\imath \infty} \frac{d \omega}{2 \pi \imath}
 \left(w/\mu^2\right)^{\omega} f^{(A)}_{q,g} \left(\omega, Q^2/\mu^2\right).
\end{eqnarray}
Throughout the paper we will address  $f^{(A)}_{q,g}$ as the Mellin amplitudes.
In what follows we will use the logarithmic variables $\rho$ and $y$:

 \begin{equation}\label{rhoy}
\rho = \ln (w/\mu^2), \xi = \ln (w/Q^2), y = \ln \left(Q^2/\mu^2\right)
\end{equation}

The transform inverse to (\ref{mellin}) is

 \begin{equation}\label{invmellin}
 f^{(A)}_{q,g} \left(\omega, Q^2/\mu^2\right) = \frac{1}{\omega} \int_{\mu^2}^{\infty} \frac{d w}{w} \left(w/\mu^2\right)^{- \omega}~A_{q,g} (w, Q^2).
 \end{equation}

 The same form of transforms (\ref{mellin},\ref{invmellin}) we will use for amplitudes $B_{q,g}$, denoting their Mellin amplitudes $f^{(B)}_{q,g}$.

\subsection{Constructing IREEs for amplitudes $A_{q,g}$ and $B_{}q,g$}

We start by constructing  IREEs for $A_{q,g}$.
The l.h.s of each IREE is obtained with differentiation of $A_{q,g}$ over $\mu$. It follows from Eq.~(\ref{mellin}) that

\begin{equation}\label{lhs}
 - \mu^2 d A_{q,g} / d \mu^2 = \partial A_{q,g}/ \partial \rho + \partial A_{q,g}/ \partial y =
 \int_{- \imath \infty}^{\imath \infty} \frac{d \omega}{2 \pi \imath}
 \left(w/\mu^2\right)^{\omega} \left[\omega + \partial f^{(A)}_{q,g} / \partial y \right].
\end{equation}

Eq.~(\ref{lhs}) represents the l.h.s. of the IREEs for $A_{q,g}$.
The guiding
principle to obtain the r.h.s. is to look for a $t$-channel parton with minimal transverse momentum
$\equiv k_{\perp}$, so $\mu$ is the lowest limit of integration over $k_{\perp}$.
Integration over $k_{\perp}$ yields a DL contribution only when there is a two-parton
intermediate state in the $t$-channel. Such pairs can consist of quarks or gluons. They factorize
 $A_{q,g}$ into two amplitudes.
Applying to them the operator $-\mu^2 \partial/\partial \mu^2$  leads to the following IREEs:

  \begin{eqnarray}\label{eqfyf}
\left[\partial/\partial y + \omega \right] f^{(A)}_{q}(\omega,y) &=&
1/(8 \pi^2)
f^{(A)}_{q} (\omega, y) f_{qq} (\omega) + 1/(8 \pi^2) f^{(A)}_{g} (\omega, y) f_{gq} (\omega),
\\ \nonumber
\left[\partial/\partial y + \omega \right] f^{(A)}_{g}(\omega, y) &=&
1/(8 \pi^2)
f^{(A)}_{q} (\omega, y) f_{qg} (\omega) + 1/(8 \pi^2) f^{(A)}_{g} (\omega, y) f_{gg} (\omega),
\end{eqnarray}
where amplitudes $f_{rr^{\prime}}$ ($r, r^{\prime} = q,g$) are the parton-parton amplitudes.
In order to get rid of the factors $1/(8 \pi^2)$ in Eq.~(\ref{eqfyf}) we replace $f_{rr^{\prime}}$
by

\begin{equation}\label{hik}
h_{rr^{\prime}} =  f_{rr^{\prime}}/(8 \pi^2)
\end{equation}
and rewrite (\ref{eqfyf}) in the following way:

  \begin{eqnarray}\label{eqfy}
\partial f^{(A)}_{q} (\omega, y)/\partial y &=&
\left[- \omega + h_{qq}(\omega)\right] f^{(A)}_{q}(\omega, y)    +  f^{(A)}_{g} (\omega, y) h_{gq} (\omega),
\\ \nonumber
\partial f^{(A)}_{g}(\omega, y)/\partial y   &=&
f^{(A)}_{q} (\omega, y) h_{qg} (\omega) + \left[- \omega +  h_{gg} (\omega)\right] h^{(A)}_{g} (\omega, y).
\end{eqnarray}

Eq.~(\ref{eqfy})
looks quite similarly to the DGLAP equations\cite{dglap}, with  $h_{rr^{\prime}}$ being new
anomalous dimensions. They accommodate double-logarithmic (DL) contributions to all orders in $\alpha_s$ and
can be calculated in DLA with applying the same method: constructing and solving appropriate IREEs for them.
The DL contributions in the $n^{th}$ order are $\sim \alpha^n_s \ln^{2n} s$, i.e. in the Mellin space they are
 $\sim \alpha^n_s/\omega^{1 + 2n}$. They are the most singular
terms at $\omega = 0$, i.e. at small $x$, so total resummation of them is important for generalizing DGLAP to the small-$x$ region.
In order to specify a  general solution to Eq.~(\ref{eqfy}) we use the matching

\begin{equation}\label{match}
A_q (\rho, y)|_{y = 0} = \widetilde{A}_q (\rho),~~~A_g (\rho, y)|_{y = 0} = \widetilde{A}_g (\rho),
\end{equation}
where $\widetilde{A}_q$ and $\widetilde{A}_g$ are the amplitudes of the same process at $Q^2 \sim \mu^2$. We denote
 $\widetilde{f}^{(A)}_q$ and $\widetilde{f}^{(A)}_g$ the Mellin amplitudes conjugated to them.
 Amplitudes $\widetilde{f}^{(A)}_{q,g}$  should be found
independently. The IREEs for them do not contain derivatives because $Q^2 = \mu^2$, so they are algebraic equations but
in contrast to Eqs.~(\ref{eqfyf},\ref{eqfy}) they are inhomogeneous:

  \begin{eqnarray}\label{eqfa}
\omega \widetilde{f}^{(A)}_{q}(\omega) &=&
\phi^{(A)}_{q} +
\widetilde{f}^{(A)}_{q} (\omega) h_{qq} (\omega) + \widetilde{f}^{(A)}_{g} (\omega) h_{gq} (\omega),
\\ \nonumber
\omega \widetilde{f}^{(A)}_{g}(\omega) &=&
\phi^{(A)}_{g} +
\widetilde{f}^{(A)}_{q} (\omega) h_{qg} (\omega) +  \widetilde{f}^{(A)}_{g} (\omega) h_{gg} (\omega),
\end{eqnarray}
with inhomogeneous terms $\phi^{(A)}_{g}$ and $\phi^{(A)}_{g}$. We will call $\phi^{(A)}_{q,g}$ inputs.
We will specify them later.
The technology of composing IREEs for amplitudes $B_q$ and $B_g$ is absolutely the same. As a result, the equations for
 $f^{(B)}_q$ and $f^{(B)}_g$ in the region $Q^2 \gg \mu^2$ are

  \begin{eqnarray}\label{eqby}
\partial f^{(B)}_{q} (\omega, y)/\partial y &=&
\left[- \omega + h_{qq}(\omega)\right] f^{(B)}_{q}(\omega, y)    +  f^{(B)}_{g} (\omega, y) h_{gq} (\omega),
\\ \nonumber
\partial f^{(B)}_{g}(\omega, y)/\partial y   &=&
f^{(B)}_{q} (\omega, y) h_{qg} (\omega) + \left[- \omega +  h_{gg} (\omega)\right] f^{(B)}_{g} (\omega, y)
\end{eqnarray}
while $f^{(B)}_{q,g}$ at  $Q^2 \sim \mu^2$ obey the following IREEs:

  \begin{eqnarray}\label{eqfb}
\omega \widetilde{f}^{(B)}_{q}(\omega) &=&
\phi^{(B)}_{q} +
\widetilde{f}^{(B)}_{q} (\omega) h_{qq} (\omega) +  \widetilde{f}^{(B)}_{g} (\omega) h_{gq} (\omega),
\\ \nonumber
\omega \widetilde{f}^{(B)}_{g}(\omega) &=&
\phi^{(B)}_{g} +
\widetilde{f}^{(B)}_{q} (\omega) h_{qg} (\omega) +  \widetilde{f}^{(B)}_{g} (\omega) h_{gg} (\omega),
\end{eqnarray}
where  $\phi^{(B)}_{g}$ and $\phi^{(B)}_{g}$ are the inputs. They differ from the inputs $\phi^{(A)}_{q,g}$ for amplitudes
$A_{q,g}$.
We will specify them later.

\section{Solution to IREEs for amplitudes $A_q$ and $A_g$}

By obtaining first general solutions to differential equations in Eqs.~(\ref{eqfy}), then solving
algebraic equations (\ref{eqfa}) and using the matching (\ref{match}) to specify  the general solutions, we
arrive\footnote{Details can be found in Ref.~\cite{etf1}.}
at the following expressions for $A_q$ and $A_g$:

\begin{eqnarray}\label{aqg}
A_q &=&  \int_{- \imath \infty}^{\imath \infty} \frac{d \omega}{2 \pi \imath} x^{- \omega}
\left[\phi^{(A)}_{q} \left( C^{(+)}_q e^{\Omega_{(+)} y} + C^{(-)}_q e^{\Omega_{(-)} y}\right) +
\phi^{(A)}_{g} \left( C^{(+)}_g e^{\Omega_{(+)} y} + C^{(-)}_g e^{\Omega_{(-)} y}\right)
\right],
\\ \nonumber
A_g &=&  \int_{- \imath \infty}^{\imath \infty} \frac{d \omega}{2 \pi \imath} x^{- \omega}
\left[\phi^{(A)}_{q} \left( \widetilde{C}^{(+)}_q e^{\Omega_{(+)} y} + \widetilde{C}^{(-)}_q e^{\Omega_{(-)} y}\right) +
\phi^{(A)}_{g} \left( \widetilde{C}^{(+)}_g e^{\Omega_{(+)} y} + \widetilde{C}^{(-)}_g e^{\Omega_{(-)} y}\right)
\right],
\end{eqnarray}
where the anomalous dimensions $\Omega_{(\pm)}$ are made of $h_{rr^{\prime}}$:

\begin{equation}\label{omegapm}
\Omega_{(\pm)} = \frac{1}{2} \left[ h_{gg} + h_{qq} \pm \sqrt{R}\right],
\end{equation}
with

\begin{equation}\label{r}
R = (h_{gg} + h_{qq})^2 - 4(h_{qq}h_{gg} - h_{qg}h_{gq}) = (h_{gg} - h_{qq})^2  + 4 h_{qg}h_{gq}.
\end{equation}

Coefficient functions $C^{(\pm)}_{q,g} (\omega) $ and $\widetilde{C}^{(\pm)}_{q,g} (\omega)$ are
also made of $h_{rr^{\prime}}$. However,
explicit expressions for them are rather bulky, so we put them in Appendix.
The IREEs for amplitudes $B_{q,g}$ are quite similar to the ones for $A_{q,g}$. As a result, the expressions for them are alike
Eq.~(\ref{aqg}):

\begin{eqnarray}\label{bqg}
B_q &=&  \int_{- \imath \infty}^{\imath \infty} \frac{d \omega}{2 \pi \imath} x^{- \omega}
\left[\phi^{(B)}_{q} \left( C^{(+)}_q e^{\Omega_{(+)} y} + C^{(-)}_q e^{\Omega_{(-)} y}\right) +
\phi^{(B)}_{g} \left( C^{(+)}_g e^{\Omega_{(+)} y} + C^{(-)}_g e^{\Omega_{(-)} y}\right)
\right],
\\ \nonumber
B_g &=&  \int_{- \imath \infty}^{\imath \infty} \frac{d \omega}{2 \pi \imath} x^{- \omega}
\left[\phi^{(B)}_{q} \left( \widetilde{C}^{(+)}_q e^{\Omega_{(+)} y} + \widetilde{C}^{(-)}_q e^{\Omega_{(-)} y}\right) +
\phi^{(B)}_{g} \left( \widetilde{C}^{(+)}_g e^{\Omega_{(+)} y} + \widetilde{C}^{(-)}_g e^{\Omega_{(-)} y}\right)
\right],
\end{eqnarray}

Eqs.~(\ref{aqg}) and \ref{bqg}) involve the same coefficient functions and anomalous dimensions for $A_{q}$ and $B_{q}$
$(A_{g}$ and $B_{g})$.
The only difference between Eqs.~(\ref{aqg}) and (\ref{bqg}) is the inputs. Now we
are going to specify them.

\section{Specifying the inputs}

By definition, inputs in evolution equations stand for the starting points of the  evolution.
They  are considered elementary and cannot be obtained through evolution.

\subsection{Inputs for amplitudes $A_{q,g}$}

Evolution of  amplitude $A_q$ starts from the Born contribution which is given by the following expression:

\begin{equation}\label{aqborn}
A^{(Born)}_q = \frac{1}{\pi} g_{\mu \nu}  \Im \left[- \frac{1}{2}~\frac{\bar{u}(p)\gamma_{\nu}\left(\hat{p} + \hat{q}\right)\gamma_{\mu}
u(p)}
{w (1 - x) - \mu^2 + \imath \epsilon}\right] =
- 2 ~\delta
(1 - x - \lambda ),
\end{equation}
where we have denoted $\lambda = \mu^2/w$. In Eq.~(\ref{aqborn}) we
dropped the quark electric charge and introduced the IR cut-off $\mu$.
It is clear that $\lambda$ can be neglected compared to $x$ in the kinematics $Q^2 \gg \mu^2$, so $A^{(Born)}_q $ is
$\mu$-independent in this region and vanishes after differentiation over $\mu$.
It explains why the quark inhomogeneous term
 is absent in Eqs.~(\ref{eqfyf},\ref{eqfy}).
In contrast,
$A^{(Born)}_q $ depends on
$\mu$
in the region $Q^2 \sim \mu^2$.  and appears in Eq.~(\ref{eqfa}).
 Dropping $x$ in Eq.~(\ref{aqborn}) compared to $\lambda$ in this region and applying the transform of Eq.~(\ref{invmellin}),
 we obtain the input $\phi^{(A)}_{q}$.  Remembering that there is no Born
 contribution to $A_g$ at any $Q^2$, we arrive at the following expressions for the inputs:

  \begin{eqnarray}\label{phia}
  \phi^{(A)}_{q} &=& - 2 ,
  \\ \nonumber
\phi^{(A)}_{g} &=& 0.
\end{eqnarray}

\subsection{Inputs for amplitudes $B_{q,g}$}

The situation with specifying inputs $\phi^{(B)}_{q,g}$ is more involved. In the Born approximation we have

\begin{equation}\label{bqborn}
B^{(Born)}_q = \frac{1}{\pi} \frac{p_{\mu} p_{\nu}}{pq}  \Im \left[- \frac{1}{2}~\frac{\bar{u}(p)\gamma_{\nu}\left(\hat{p} + \hat{q}\right)\gamma_{\mu}
u(p)}
{w (1 - x) - \mu^2 + \imath \epsilon}\right] \sim p^2 \approx 0.
\end{equation}

In addition, $B^{(Born)}_g = 0$.  Thus, the both Born amplitudes are zeros, which excludes using the Born
approximation as the starting point of the evolution. They are non-zeros in the first loop:

\begin{equation}\label{b1}
B_q^{(1)} = \frac{\alpha_s}{2 \pi} C_F,~~ B_q^{(1)}  =
\frac{\alpha_s}{\pi}  n_f (1-x),
\end{equation}
where we have used the standard notations $C_F = (N^2 - 1)/2N = 4/3$ and $n_f$ is the number of flavours.
Therefore, $B_{q,g}^{(1)}$ could be used as inputs in IREEs for $B_{q,g}$. In this case, the result of
evolving $B_{q,g}$ to the order $\alpha^2_s$
would  be $\sim \alpha_s^2\ln^{2} (w/\mu^2)$. However, the most important
second-loop contributions to $B_{q,g}$ are proportional to $\alpha^2_s/x$ and they cannot be obtained
with evolving the inputs $B_{q,g}^{(1)}$.
The only way to include them in IREEs is to choose them as the inputs and evolve them with IREEs. By doing so, we skip
$B_{q,g}^{(1)}$, so they can be added by hands.  We remind that calculations of $F_{1,2}$ in
the first and second loops can be found in Refs.~\cite{flor} - \cite{kotik} whereas Ref.~\cite{moch3}
contains the third-loop calculations.  In the present paper we will use the leading second-loop contributions
$B_q^{(2)} (B_g^{(2)})$
to $B_q (B_g)$ obtained in Ref.~\cite{etfl}. They are given by the following expressions:

\begin{eqnarray}\label{b2qglead}
B^{(2)}_q &=& C^{(2)}_q  \gamma^{(2)} \rho~x^{-1},
\\ \nonumber
B^{(2)}_g &=& C^{(2)}_g  \gamma^{(2)} \rho~x^{-1},
\end{eqnarray}
with

\begin{equation}\label{gamma2}
\gamma^{(2)} = \alpha^2_s \ln 2/2 \pi
\end{equation}
and  the color factors
$C^{(2)}_q = C_F^2 + C_F n_f$ and $C^{(2)}_g = N n_f + C_F n_f$ . We use $B_{q,g}^{(2)}$ as the starting point of IR evolution.
Combining Eqs.~(\ref{invmellin}) and (\ref{b2qglead}), we obtain the following expressions for the inputs $\phi^{(B)}_{q,g}$:

  \begin{eqnarray}\label{phib}
  \phi^{(B)}_{q} &=&  x^{-1} \gamma^{(2)} \rho  ~C^{(2)}_q,
  \\ \nonumber
\phi^{(B)}_{g} &=& x^{-1}  \gamma^{(2)} \rho  ~C^{(2)}_g.
\end{eqnarray}

The factor $\rho$  in Eq.~(\ref{b2qglead}) is not replaced by $1/\omega^2$ in Eq.~(\ref{phib}) in
contrast  to Ref.~\cite{etfl}
by the following reason: $\rho$ was not obtained through the IR evolution,
so it does not participate in the Mellin transform. This point is the main difference between the present paper
and our previous paper Ref.~\cite{etfl}.

\section{Expressions for $F_1$ and $F_2$}

Using Eqs.~(\ref{phia},\ref{phib}) to specify inputs in Eqs.~(\ref{aqg},\ref{bqg}), we obtain explicit expressions for
$A_{q,g}$ and $B_{q,g}$. Combining these results with Eq.~(\ref{fab}), we arrive at explicit expressions for
$F_1^{(q,g)}$ and $F_2^{(q,g)}$:

\begin{eqnarray}\label{f1qgi}
F_1^{(q)} &=&    \left(1 + \gamma^{(2)}\rho C^{(2)}_q\right) I_q
+ \gamma^{(2)}\rho C^{(2)}_g  I_g ,
\\ \nonumber
F_1^{(g)} &=& \left(1 + \gamma^{(2)} \rho C^{(2)}_q\right) \widetilde{I}_g
+ \gamma^{(2)}\rho  C^{(2)}_g  \widetilde{I}_g,
\end{eqnarray}

\begin{eqnarray}\label{f2qgi}
F_2^{(q)} &=&    2x \left[\left(1 + 3 \gamma^{(2)}\rho3  C^{(2)}_q\right) I_q
+ 3 \gamma^{(2)} \rho C^{(2)}_g \rho I_g \right],
\\ \nonumber
F_2^{(g)} &=& 2x \left[ \left(1 + 3\gamma^{(2)} \rho C^{(2)}_q\right) \widetilde{I}_q
+ 3 \gamma^{(2)} \rho C^{(2)}_g \widetilde{I}_g \right],
\end{eqnarray}
with $I_{q,g}$ and $\widetilde{J}_{q,g}$ being defined as follows:

\begin{eqnarray}\label{iqg}
I_q &=& \int_{- \imath \infty}^{\imath \infty} \frac{d \omega}{2 \pi \imath} x^{- \omega}
\left(C^{(+)}_q  e^{\Omega_{(+)} y} +  C^{(-)}_q e^{\Omega_{(-)} y}\right),
\\ \nonumber
I_g &=& \int_{- \imath \infty}^{\imath \infty} \frac{d \omega}{2 \pi \imath} x^{- \omega}
 \left(C^{(+)}_g  e^{\Omega_{(+)} y} +  C^{(-)}_g e^{\Omega_{(-)} y}\right),
\\ \nonumber
\widetilde{I}_q &=&  \int_{- \imath \infty}^{\imath \infty} \frac{d \omega}{2 \pi \imath} x^{- \omega}
  \left(\widetilde{C}^{(+)}_q  e^{\Omega_{(+)} y} +  \widetilde{C}^{(-)}_q e^{\Omega_{(-)} y}\right),
\\ \nonumber
\widetilde{I}_g &=&  \int_{- \imath \infty}^{\imath \infty} \frac{d \omega}{2 \pi \imath} x^{- \omega}
 \left(\widetilde{C}^{(+)}_g  e^{\Omega_{(+)} y} +  \widetilde{C}^{(-)}_g e^{\Omega_{(-)} y}\right).
\end{eqnarray}

Comparison of Eqs.~(\ref{f1qgi}) and (\ref{f2qgi}) yields that $F_L^{(q,g)}$ are given by the following expressions

\begin{eqnarray}\label{flqgi}
F_L^{(q)} &=&  4x \gamma^{(2)}\rho \left[ C^{(2)}_q I_q + C^{(2)}_g I_g \right],
\\ \nonumber
F_L^{(g)} &=& 4x \gamma^{(2)} \rho \left[ C^{(2)}_q \widetilde{I}_q + C^{(2)}_g \widetilde{I}_g \right].
\end{eqnarray}

We remind that explicit expressions for $C^{(\pm)}_{q}, C^{(\pm)}_{g}, \widetilde{C}^{(\pm)}_{q}$
and  $\widetilde{C}^{(\pm)}_{g} $ can be found in Appendix A.
Eqs.~(\ref{f1qgi}, \ref{f2qgi},\ref{flqgi}) can be interpreted as a generalization of the DGLAP expressions for
the structure functions $F^{(q,g)}_1,F^{(q,g)}_2,F^{(q,g)}_L $ on the case of small $x$.
We would like to mention that in our previous paper\cite{etf1} on the structure function $F_1$ we
neglected the contributions $\sim \gamma^{(2)}$ in Eq.~(\ref{f1qgi}). Fortunately, this flaw, being essential by itself,
does not
change qualitatively intercept of the Reggeon which controls the small-$x$ asymptotics of $F_1$ as well as the
$Q^2$-dependence of $F_1$, which were the main objectives of Ref.~\cite{etf1}.
Eqs.~(\ref{f1qgi},\ref{f2qgi},\ref{flqgi}) are valid at $Q^2 > \mu^2$  ( $\mu \approx 1$~GeV, see
Ref.~\cite{egtg1sum} for detail)
but it is easy to
generalize them for smaller $Q^2$. The prescription is obtained in Ref.~\cite{etcomb}:
Eqs.~(\ref{f1qgi},\ref{f2qgi},\ref{flqgi}) can be used at arbitrary $Q^2$ providing that $Q^2$ is replaced
by $\bar{Q}^2 = Q^2 + \mu^2$. It leads to replacing $x,y$ and $\xi$ by $\bar{x}, \bar{y}, \bar{\xi}$ respectively:

\begin{equation}\label{shift}
\bar{x} = \left(Q^2 + \mu^2\right)/w, ~\bar{y} = \left(Q^2 + \mu^2\right)/\mu^2, ~\bar{\xi} =
\ln \left[\left(Q^2 + \mu^2\right)/w \right].
\end{equation}

Convoluting Eqs.~(\ref{f1qgi}, \ref{f2qgi},\ref{flqgi}) with the parton distributions
$\Phi_q$ and $\Phi_g$, we obtain expressions for the
structure functions $F_1,F_2,F_L$ of the unpolarized DIS, which can be used at small $x$ and arbitrary $Q^2$:

\begin{eqnarray}\label{f12l}
F_1 &=&    \left(1 + \gamma^{(2)} \rho~ C^{(2)}_q\right) \left(J_q + \widetilde{J}_q \right)
+ \gamma^{(2)} \rho~ C^{(2)}_g  \left(J_g + \widetilde{J}_g \right) ,
\\ \nonumber
F_2  &=& 2x \left[ \left(1 + 3\gamma^{(2)} \rho~ C^{(2)}_q\right) \left(J_q + \widetilde{J}_q\right)
+ 3 \gamma^{(2)}\rho~ C^{(2)}_g   \left(J_g + \widetilde{J}_g\right) \right],
\\ \nonumber
F_L &=& 4x \gamma^{(2)} \rho \left[ C^{(2)}_q \left(J_q + \widetilde{J}_q \right)
 + C^{(2)}_g  \left(J_g + \widetilde{J}_g\right) \right].
\end{eqnarray}

Each of $J_{q,g}$ and $\widetilde{J}_{q,g}$ includes both the perturbative integrands of Eq.~(\ref{iqg}) and the
non-perturbative parton distributions $\widetilde{\Phi}_{q,g}(\omega)$:

\begin{eqnarray}\label{jqg}
J_q &=& \int_{- \imath \infty}^{\imath \infty} \frac{d \omega}{2 \pi \imath} \bar{x}^{- \omega}
\left(C^{(+)}_q  e^{\Omega_{(+)} \bar{y}} +  C^{(-)}_q e^{\Omega_{(-)} \bar{y}}\right)
\hat{\Phi}_q (\omega),
\\ \nonumber
J_g &=& \int_{- \imath \infty}^{\imath \infty} \frac{d \omega}{2 \pi \imath} \bar{x}^{- \omega}
 \left(C^{(+)}_g  e^{\Omega_{(+)} \bar{y}} +  C^{(-)}_g e^{\Omega_{(-)} \bar{y}}\right)
 \hat{\Phi}_g (\omega),
\\ \nonumber
\bar{J}_q &=&  \int_{- \imath \infty}^{\imath \infty} \frac{d \omega}{2 \pi \imath} \bar{x}^{- \omega}
  \left(\widetilde{C}^{(+)}_q  e^{\Omega_{(+)} \bar{y}} +  \widetilde{C}^{(-)}_q e^{\Omega_{(-)} \bar{y}}\right)
  \hat{\Phi}_q (\omega),
\\ \nonumber
\widetilde{J}_g &=&  \int_{- \imath \infty}^{\imath \infty} \frac{d \omega}{2 \pi \imath} \bar{x}^{- \omega}
 \left(\widetilde{C}^{(+)}_g  e^{\Omega_{(+)} \bar{y}} +  \widetilde{C}^{(-)}_g e^{\Omega_{(-)} \bar{y}}\right)
 \hat{\Phi}_g (\omega).
\end{eqnarray}

The notations $\hat{\Phi}_{q,g} (\omega)$ stand for the parton distributions in the $\omega$-space. They are
related to the parton distributions $\Phi_{q,g}$ in the $x$-space by the transform (\ref{mellin}).
At this point we stress once more that, in contrast to DGLAP, the parton distributions $\Phi_{q,g}$
 should not involve factors $x^{-a}$. The role of such factors is to ensure the structure functions
 with fast growth at $x \to 0$. However, the
total resummation of DL contributions in Eq.~(\ref{iqg}) automatically leads to the Regge asymptotics at small $x$,
which we will demonstrate in the
next Sect.

\section{Small-$x$ asymptotics of the unpolarized structure functions}

Eq.~(\ref{omegapm}) reads that $\Omega_{(+)} > \Omega_{(-)}$. Because of that we drop the terms comprising  $\Omega_{(-)}$
in Eq.~(\ref{jqg}), when calculate the small-$x$ asymptotics of $J_{q,g}$ and $\widetilde{J}_{q,g}$.
 Then we represent Eq.~(\ref{jqg}) in the following exponential
form:

\begin{eqnarray}\label{jqgpsi}
J_{q,g} &\approx&  \int_{- \imath \infty}^{\imath \infty} \frac{d \omega}{2 \pi \imath}
e^{\omega \bar{\xi}  + \Psi_{q,g}(\omega) },
\\ \nonumber
\widetilde{J}_{q,g} &\approx&  \int_{- \imath \infty}^{\imath \infty} \frac{d \omega}{2 \pi \imath}
e^{\omega \bar{\xi}  + \widetilde{\Psi}_{q,g}(\omega) },
\end{eqnarray}
with

\begin{eqnarray}\label{f12lpsi}
 \Psi_{q,g} &=& \Omega_{(+)} \bar{y} + \ln C^{(+)}_{q,g},
 \\ \nonumber
\widetilde{\Psi}_{q,g} &=& \Omega_{(+)} \bar{y}  + \ln \widetilde{C}^{(+)}_{q,g}.
\end{eqnarray}

Now we push $\bar{x} \to 0$ (i.e. $\xi \to \infty$) and apply the Saddle-Point method to each expression in
Eq.~(\ref{jqgpsi}). Handling any  of $J_{q,g}$ and $\widetilde{J}_{q,g}$ is the same.
This method states that the small-$x$ asymptotics of the structure functions is given by the following expressions
(see Appendix C for detail):

\begin{eqnarray}\label{asf12l}
F_1 &\sim&    \frac{\chi_1}{\xi^{3/2}}
~\bar{x}^{- \omega_0} \left(\bar{Q}^2/\mu^2\right)^{\omega_0/2} ,
\\ \nonumber
F_2  &\sim& \frac{\chi_2}{\xi^{3/2}}~\bar{x}^{- \omega_0 +1} \left(\bar{Q}^2/\mu^2\right)^{\omega_0/2},
\\ \nonumber
F_L   &\sim& \frac{\chi_L}{\xi^{3/2}}~\bar{x}^{- \omega_0 +1} \left(\bar{Q}^2/\mu^2\right)^{\omega_0/2},
\end{eqnarray}
where $\bar{x}$ and $\bar{Q}^2$ are defined in Eq.~(\ref{shift}), so Eq.~(\ref{asf12l})
is valid for large and small $Q^2$, including $Q^2 = 0$. The non-exponential factors $\chi_{1,2,L} (\omega_0)$
are defined as follows:

\begin{eqnarray}\label{chi}
\chi_1 &=& \left(1 + \gamma^{(2)} \rho~ C^{(2)}_q\right) C^{(+)}_q (\omega_0) \lambda_q (\omega_0)
+ \gamma^{(2)} \rho~ C^{(2)}_g  C^{(+)}_g (\omega_0) \lambda_g (\omega_0),
\\ \nonumber
\chi_2 &=& 2 \left(1 + 3 \gamma^{(2)} \rho~ C^{(2)}_q\right) C^{(+)}_q (\omega_0) \lambda_q (\omega_0)
+ 3 \gamma^{(2)} \rho~ C^{(2)}_g  C^{(+)}_g (\omega_0) \lambda_g (\omega_0),
\\ \nonumber
\chi_L &=& 4 \gamma^{(2)} \rho~\left(C^{(2)}_q C^{(+)}_q (\omega_0) \lambda_q (\omega_0)
+ C^{(2)}_g  C^{(+)}_g (\omega_0) \lambda_g (\omega_0)\right).
\end{eqnarray}

Obviously,

\begin{equation}\label{qbig}
\left(\bar{Q}^2/\mu^2\right)^{\omega_0/2} \approx \left(Q^2/\mu^2\right)^{\omega_0/2}
\end{equation}
 at $Q^2 \gg \mu^2$ (we remind that $\mu \approx 1$~GeV, see for detail Ref.~\cite{egtg1sum})
while in the region of small $Q^2$, at $Q^2 \ll \mu^2$,

\begin{equation}\label{qsmall}
\left(\bar{Q}^2/\mu^2\right)^{\omega_0/2} \approx 1 + \left(\omega_0/2\right) \left(Q^2/\mu^2\right).
\end{equation}

The notation $\omega_0$ in Eq.~(\ref{asf12l})
stands for the rightmost singularities of the perturbative factors $\Psi_{q,g}$ and $\widetilde{\Psi}_{q,g}$  in Eq.~(\ref{f12lpsi}). It turns out that
the rightmost singularity of any them is the rightmost root of the equation $W = 0$, with $W$ being defined in Eq.~(\ref{w}).
Explicitly, this equation is

\begin{equation}\label{omegaeq}
(\omega^2 - 2(b_{qq} + b_{gg}))^2 - 4 (b_{qq} - b_{gg})^2 - 16b_{gq} b_{qg} = 0,
\end{equation}
with $b_{ik}$ defined in Eq.~(\ref{bik}). As $W$ takes place in expressions for each structure function,
$\omega_0$ is the same for any of them.  The factors $\chi_{1,2,3}$ include as numerical factors of the perturbative origin as
the parton distributions $\hat{\Phi}_{q,g} (\omega)$ at $\omega = \omega_0$. Those factors are different for different structure functions.
Eq.~(\ref{omegaeq}) can be solved analytically for the fixed $\alpha_s$ approximation only\footnote{see Ref.~\cite{etf1} for detail}.
When $\alpha_s$ runs, Eq.~(\ref{omegaeq}) has to be solved numerically, which leads to the value

\begin{equation}\label{omegazero}
\omega_0 = 1.07.
\end{equation}

In contrast to the $x$-dependence, the asymptotics of $F_{1,2,L}$ in Eq.~(\ref{asf12l}) exhibit  the almost
identical $Q^2$-dependence. It is generated by the term $e^{\Omega_{(+)}(\omega) y}$ which
participates in each $\Psi_r$.
Using explicit expressions for $h_{rr^{\prime}}$ in Appendix B, it is easy to show
that $\Omega_{(+)}(\omega_0) = \omega_0/2$.

Eq.~(\ref{asf12l}) demonstrates that asymptotics of all structure functions are of the Regge type.
The Saddle-Point method turns the total sum of the terms $\sim \left(\alpha_s \ln^2 (1/x)\right)^n$ into
the Regge power factor $x^{- \omega_0}$. It grows steeply at small $x$, which makes redundant
factors $x^{-a}$ in the parton distributions $\Phi_{q,g}$.
The intercept of the Reggeon controlling $F_1$ exceeds unity, so it is a new (soft) Pomeron. Although it has nothing in common
with the BFKL Pomeron, its intercept is surprisingly close to the one of the BFKL Pomeron in NLO.
This issue  was considered in detail in Ref.~\cite{etf1}.   In contrast, the intercepts of the other Reggeons in Eq.~(\ref{asf12l})
are much smaller than unity but nevertheless they predict the slow growth of $F_2$ and $F_L$
when $x$ decreases.
 Below we briefly
consider some corollaries of Eq.~(\ref{asf12l}).

\subsection{Asymptotic scaling}

The asymptotics in Eq.~(\ref{asf12l}) at $Q^2 \gg \mu^2$ can approximately be represented as follows:

\begin{eqnarray}\label{scaling}
F_1 &\sim&
\zeta^{-1.07},
\\ \nonumber
F_2 &\sim&
x \zeta^{-1.07},
\\ \nonumber
F_L &\sim&
x \zeta^{-1.07},
\end{eqnarray}
with $\zeta = x\sqrt{\mu^2/Q^2}$, so that $F_1$ as well as $F_2/x$ and $F_L/x$ at $x \ll 1$ and $Q^2 \gg \mu^2$
depend on the argument $\zeta$ save the logarithmic factors dropped in Eq.~(\ref{scaling}). Such scaling of the asymptotics of the
structure functions has
not been predicted by any other approach.

\subsection{Ratio $F_L/F_2$ at small $x$}

It follows from Eq.~(\ref{asf12l}) that the ratio $F_L/F_2$ is given by the following expression:

\begin{equation}\label{ratio}
R_{2L} \equiv F_L/F_2 = \frac{2 \gamma^{(2)} \rho~\left(C^{(2)}_q C^{(+)}_q (\omega_0) \lambda_q (\omega_0)
+ C^{(2)}_g  C^{(+)}_g (\omega_0) \lambda_g (\omega_0)\right)}
{\left(1 + 3 \gamma^{(2)} \rho~ C^{(2)}_q\right) C^{(+)}_q (\omega_0) \lambda_q (\omega_0)
+ 3 \gamma^{(2)} \rho~ C^{(2)}_g  C^{(+)}_g (\omega_0) \lambda_g (\omega_0)}.
\end{equation}

We remind that $\gamma^{(2)} $ is defined in Eq.~(\ref{gamma2}) and $\rho = \xi + y \approx \xi$ at $x \ll 1$.
Obviously, $F_L/F_2 \approx \gamma^{(2)} \rho \approx 0.006 \rho$ at $ \rho \ll 1/\gamma^{(2)}$ which corresponds to the energy scale 
presently available at experiment.
In the opposite case, i.e. at $ \rho \gg 1/\gamma^{(2)}$, the ratio
$F_L/F_2 \sim 2/3$, though this limit can be achieved  at really asymptotic energies.

\subsection{Relations between logarithmic derivatives of the structure functions}

Logarithmic derivatives, i.e. $\partial \ln F_r/ \partial \ln Q^2 = (1/F_r) \partial F_r/ \partial \ln Q^2$, with $r = 1,2,L$, 
were already discussed in the literature in the context of DGLAP and the dipole model
(see e.g. Refs.~\cite{kotik2, bor}).
It motivates us to construct analogous relations for $F_r$ in DLA at the small-$x$ by differentiating Eq.~(\ref{asf12l}).
 First of all, there are relations for the $Q^2$-dependence of  the
structure functions:

\begin{equation}\label{logq}
\frac{\partial \ln F_1}{\partial y} = \frac{\partial \ln F_2}{\partial y} \approx \frac{\partial \ln F_L}{\partial y}.
\end{equation}

Then, the relations involving the $x$- and $Q^2$-dependence of  $F_r$:

\begin{eqnarray}\label{logxq}
\frac{\partial \ln F_1}{\partial \xi} - 2 \frac{\partial \ln F_1}{\partial y} &\sim& 0,
\\ \nonumber
\frac{\partial \ln F_2}{\partial \xi} - 2 \frac{\partial \ln F_2}{\partial y} &\sim& - 1,
\\ \nonumber
\frac{\partial \ln F_L}{\partial \xi} - 2 \frac{\partial \ln F_L}{\partial y} &\sim& - 1
\end{eqnarray}
at $x \ll 1$ and $Q^2 > \mu^2 \approx 1$~GeV$^2$.  We stress that the relations
(\ref{logxq}) differ a lot from the results
in all approaches based on BFKL and DGLAP or on their modifications (see e.g. Refs.~\cite{kovch,kotik2}).
The difference between our results and the results (see e.g. Ref.~\cite{duc})
obtained in the Regge inspired models\cite{cap,bart} is even greater: the intercepts in
our approach do not depend on $Q^2$.

\subsection{Remark on Soft and Hard Pomerons}

One of results obtained in Ref.~\cite{etf1} is the estimate of the region of applicability of the Regge
asymptotics: the expressions for the small-{x} asymptotics
in Eq.~(\ref{asf12l}) are
reliable at $x \leq 10^{-6}$. The straightforward way to describe $F_{1,2}$
and $F_L$ at lager $x$ is to apply the parent expressions of Eqs.~(\ref{f12l})
despite their complexity. The same should be done, when BFKL is applied. However, there is a tendency in the literature
to use
the Regge asymptotics at $x \gg 10^{-6}$, which inevitably leads to introducing phenomenological Pomerons with
intercepts much greater than $1.07$. In order to simplify our explanation we
use the generic notation $F$ for any of $F_{1,2}, F_L$ and denote $\Gamma$ their small-$x$ asymptotics.
The ratio

\begin{equation}\label{rasdef}
R_{as} \equiv \Gamma/F
\end{equation}
depends on both $Q^2$ and $x$, i.e. $R_{as}= R_{as}(x,Q^2)$. To begin with, we put $Q^2 = \mu^2$ and study dependence of
$R_{as} (x,\mu^2)$ on $x$. It turns out that
at $x_0 = 10^{-6}, x_1 =10^{-4}, x_2 = 10^{-3}$ the values of $R_{as} (x,\mu^2)$ are as follows:

\begin{eqnarray}\label{ras}
R_{as} \left(x_0,\mu^2\right) &=& 0.9,
\\ \nonumber
R_{as} \left(x_1,\mu^2\right) &=& 0.67,
\\ \nonumber
R_{as} \left(x_2,\mu^2\right) &=& 0.5.
\end{eqnarray}

Eq.~(\ref{ras}) explains why $x_0$ was chosen in Ref.~\cite{etf1} as the border of applicability region of the asymptotics.
Numerical estimates show that $R_{as}$  decreases when $Q^2$ grows.
On the other hand, in practice the Regge asymptotics are used at $x > x_0$. In this case
a new Reggeon is supposed to mimic the structure functions and as a result it should equate the Regge factor of
Eq.~(\ref{asf12l}):

\begin{equation}\label{eq}
x_0^{-\omega_0} = x^{-a}.
\end{equation}

For example, if the asymptotics is used at $x_1 =10^{-4}$,
the intercept $a_1$ of the phenomenological Reggeon is

\begin{eqnarray}\label{a12}
a_1 = \omega_0 \frac{\ln x_0}{\ln x_1} = \frac{6}{4} ~\omega_0 = 1.6.
\end{eqnarray}

This estimate demonstrates that approximating parent amplitudes by their asymptotics  beyond the applicability regions 
inevitably leads to
introducing phenomenological
hard Pomerons.

\section{Summary}

In the present paper, we have calculate the structure functions $F_1$ and $F_2$ in DLA. By doing so, we first correct
our previous results on $F_1$ obtained in Ref.~\cite{etf1} and then calculate  $F_2$.
We find that the contributions
to $F_1$ we neglected in Ref.~\cite{etf1} are sizable but do not change qualitatively basic features of $F_1$.
The instrument we use in
order to sum DL contributions to all
orders in $\alpha_s$ is the IREE method. As a result, we obtain explicit expressions for $F_1$ and $F_2$. Then
we use the Saddle-Point Method to calculate the small-$x$ asymptotics of $F_{1,2}$. These asymptotics prove to be of the
Regge type, though controlled by different Reggeons. The intercept of the Reggeon governing $F_1$ is greater than unity, so
it is a new contribution to Pomeron. In contrast, the intercept of the Reggeon governing $F_2$ is small but positive,
which leads to slow growth of $F_1$ when $x$ is decreasing. We demonstrate that DLA predicts identical $Q^2$-dependence of $F_1$ and $F_2$
and explain the reason to it. The asymptotics $F_1$ and $F_2$ are used to obtain several differential relations between logarithms of
 $F_1$ and $F_2$, which are absent in all other approaches available in the literature. It would be interesting to check these relations with
analysis of experimental data.

\section{Acknowledgements}

We are grateful to V.~Bertone, A.~Cooper-Sarkar and Yuri V.~Kovchegov for useful communications.

\section{Appendix}

\subsection{Expressions for $C^{(\pm)}_{q,g} (\omega) $ and $\widetilde{C}^{(\pm)}_{q,g} (\omega)$}

\begin{eqnarray}\label{cpm}
 C^{(+)}_q &=& \frac{(h_{qg} - \omega)\left(h_{gg} - h_{qq} - \sqrt{R}\right) + 2 h_{qg}h_{gq}}{2 \Delta \sqrt{R}},
\\ \nonumber
 C^{(-)}_q &=& \frac{-(h_{qq} - \omega)\left(h_{gg} - h_{qq} + \sqrt{R}\right) - 2 h_{qg}h_{gq}}{2 \Delta \sqrt{R}},
 \\ \nonumber
C^{(+)}_g &=&  \frac{- h_{qg}\left(h_{gg} - h_{qq} - \sqrt{R}\right) - 2 h_{qg} (h_{qq} - \omega)}{2 \Delta \sqrt{R}},
 \\ \nonumber
C^{(-)}_g &=&  \frac{h_{qg}\left(h_{gg} - h_{qq} + \sqrt{R}\right) + 2 h_{qg} (h_{qq} - \omega)}{2 \Delta \sqrt{R}}.
\end{eqnarray}

Coefficient functions $ \widetilde{C}^{(\pm)}_g$ and $ \widetilde{C}^{(\pm)}_g$ are related with $C^{(\pm)}_{q,g}$:

\begin{eqnarray}\label{ctilde}
 \widetilde{C}^{(+)}_q &=&  C^{(+)}_q  X^{(+)},~~ \widetilde{C}^{(+)}_g =  C^{(+)}_g  X^{(+)},
\\ \nonumber
\widetilde{C}^{(-)}_q &=&  C^{(-)}_q  X^{(-)},~~ \widetilde{C}^{(-)}_g =  C^{(-)}_g  X^{(-)},
\end{eqnarray}
where

\begin{equation}\label{xpm}
X^{(\pm)} = \frac{h_{gg} - h_{qq} \pm \sqrt{R}}{2h_{qg}}.
\end{equation}

Thus we have expressed all coefficient functions in Eqs.~(\ref{f1qgi} - \ref{flqgi}) through $h_{rr^{\prime}}$.

\subsection{Expressions for $h_{ik}$}

\begin{eqnarray}\label{h}
&& h_{qq} = \frac{1}{2} \Big[ \omega - Z - \frac{b_{gg} -
b_{qq}}{Z}\Big],\qquad h_{qg} = \frac{b_{qg}}{Z}~, \\ \nonumber &&
h_{gg} = \frac{1}{2} \Big[ \omega - Z + \frac{b_{gg} -
b_{qq}}{Z}\Big],\qquad h_{gq} =\frac{b_{gq}}{Z}~,
\end{eqnarray}
where
\begin{equation}
\label{z}
 Z = \frac{1}{\sqrt{2}}\sqrt{ Y + W
}~,
\end{equation}
with
\begin{equation}\label{y}
Y = \omega^2 - 2(b_{qq} + b_{gg})
\end{equation}
and
\begin{equation}\label{w}
  W = \sqrt{(\omega^2 - 2(b_{qq} + b_{gg}))^2 - 4 (b_{qq} - b_{gg})^2 -
16b_{gq} b_{qg} },
\end{equation}

where the terms $b_{rr'}$ include the Born factors $a_{rr'}$ and contributions of non-ladder graphs $V_{rr'}$:
\begin{equation}\label{bik}
b_{rr'} = a_{rr'} + V_{rr'}.
\end{equation}

The Born factors are (see Ref.~\cite{egtg1sum} for detail):

\begin{equation}\label{app}
a_{qq} = \frac{A(\omega)C_F}{2\pi},~a_{qg} = \frac{A'(\omega)C_F}{\pi},~a_{gq} = -\frac{A'(\omega)n_f}{2 \pi}.
~a_{gg} = \frac{2N A(\omega)}{\pi},
\end{equation}
where $A$ and $A'$ stand for the running QCD couplings as shown in Ref.~\cite{egtalpha}:

\begin{eqnarray}\label{a}
A = \frac{1}{b} \left[\frac{\eta}{\eta^2 + \pi^2} - \int_0^{\infty} \frac{d z e^{- \omega z}}{(z + \eta)^2 + \pi^2}\right],
A' = \frac{1}{b} \left[\frac{1}{\eta} - \int_0^{\infty} \frac{d z e^{- \omega z}}{(z + \eta)^2}\right],
\end{eqnarray}
with $\eta = \ln \left(\mu^2/\Lambda^2_{QCD}\right)$ and $b$ being the first coefficient of the Gell-Mann- Low function. When the running effects for the QCD coupling
are neglected,
$A(\omega)$ and $A'(\omega)$ are replaced by $\alpha_s$.
The terms $V_{rr'}$ approximately represent the impact of non-ladder graphs on $h_{rr'}$ (see Ref.~\cite{egtg1sum} for detail):

\begin{equation}
\label{vik} V_{rr'} = \frac{m_{rr'}}{\pi^2} D(\omega)~,
\end{equation}
with
\begin{equation}
\label{mik} m_{qq} = \frac{C_F}{2 N}~,\quad m_{gg} = - 2N^2~,\quad
m_{gq} = n_f \frac{N}{2}~,\quad m_{qg} = - N C_F~,
\end{equation}
and
\begin{equation}
\label{d} D(\omega) = \frac{1}{2 b^2} \int_{0}^{\infty} d z
e^{- \omega z} \ln \big( (z + \eta)/\eta \big) \Big[
\frac{z + \eta}{(z + \eta)^2 + \pi^2} - \frac{1}{z +
\eta}\Big]~.
\end{equation}

\subsection{Basics of Saddle-Point Method}

Any expression in Eq.~(\ref{jqg}) can be generically represented as follows:

\begin{equation}\label{sp1}
J =  \int_{- \imath \infty}^{\imath \infty} \frac{d \omega}{2 \pi \imath}
e^{\omega \xi  + \Psi(\omega) },
\end{equation}
where $\Psi$ stands for any of $\Psi_{q,g}, \widetilde{\Psi}_{q,g}$.
Let us  expand $\Psi$ in the series in vicinity of the extremum point $\omega_0$, retaining three terms only.
Then the exponent in Eq.~(\ref{sp1}) is

\begin{equation}\label{ser}
\xi \omega + \Psi (\omega) \approx \xi \omega_0 + \Psi (\omega_0) + \left[\xi + \Psi^{\prime} (\omega_0)\right] (\omega - \omega_0)
+ (1/2) \Psi^{\prime \prime} (\omega_0) (\omega - \omega_0)^2
\end{equation}
and there is extremum at $\omega = \omega_0$, then

\begin{equation}\label{extr}
\xi + \Psi^{\prime} (\omega_0) =0,
\end{equation}

Now push $\xi \to \infty$ and notice the in order to equate $\xi$ in (\ref{extr}), $\Psi^{\prime} (\omega_0)$
should be singular at
$\omega = \omega_0$. $\Psi$ has many singularities but the rightmost singularity corresponds to $W = 0$.
Then

\begin{equation}\label{psiprime}
\Psi^{\prime} (\omega_0) \approx \frac{\partial \Psi}{\partial W} \frac{d W}{d \omega}=
\left[\frac{\partial \Psi}{\partial W} \omega_0 (\omega_0 - b_{qq}- b_{gg})\right] \frac{1}{ W} \equiv \frac{\kappa_1}{W}.
\end{equation}

Combining it with Eq.~(\ref{extr}) yields

\begin{equation}\label{wxi}
W = \kappa_1/\xi.
\end{equation}

When the most singular contribution is accounted for,

\begin{equation}\label{psi2prime}
\Psi^{\prime \prime} (\omega_0) \approx
- \left[\frac{\partial \Psi}{\partial W} \omega^2_0 (\omega_0 - b_{qq}- b_{gg})^2\right] \frac{1}{ W^3} \equiv \frac{\kappa_2}{W^3}
= \frac{\kappa_2}{ \kappa^3_1}\xi^3 \equiv \lambda \xi^3
\end{equation}
and therefore

\begin{equation}\label{sp2}
J \sim  e^{\omega_0 \xi  + \Psi(\omega_0) }\int_{- \imath \infty}^{\imath \infty} \frac{d \omega}{2 \pi \imath}
e^{-(1/2) (\kappa_1/\kappa^3_2) \xi^3 (\omega - \omega_0)^2 } = e^{\omega_0 \xi  + \Psi(\omega_0)}
\sqrt{\frac{2 \pi \kappa^3_2}{\kappa_1 \xi^3}} \equiv \frac{\lambda}{\xi^{3/2}} e^{\omega_0 \xi}.
\end{equation}

\end{document}